\documentclass[a4paper]{jpconf}
\usepackage{graphicx}
\usepackage{bm}
\usepackage{color}
\begin{document}
\title{Meissner effect in the layered Kane-Mele model with Hubbard interaction}

\author{Jun Goryo and Nobuki Maeda$^1$}

\address{Institute of Industrial Science, the University of Tokyo, 4-6-1 Komaba, Tokyo, Japan\\
$^1$Department of Physics, Hokkaido University, Sapporo, Japan}

\ead{jungoryo@iis.u-tokyo.ac.jp}

\begin{abstract}
We investigate the magnetic response in the quantum spin Hall phase 
of the layered Kane-Mele model with Hubbard interaction, and argue a condition 
to obtain the Meissner effect. The effect of Rashba spin orbit coupling is also discussed. 
\end{abstract}

The Kane-Mele (KM) model is proposed to describe the quantum spin Hall effect (QSHE) of 
electrons on the two-dimensional honeycomb lattice\cite{KM1,KM2}. 
Here, we will show that, in a certain parameter region, the Meissner effect is obtained 
from the effective field theory of the layered KM model with an electronic correlation\cite{Rachel-LeHur}. 
We use $\hbar=c=1$ unit and the Minkovskian metric $g^{\mu\nu}=diag(1,-1,-1)$, where 
$\mu,\nu=0,x,y$. 
  

One of the essential ingredients of the KM model\cite{KM1,KM2} is the intrinsic spin-orbit (SO) coupling 
$\lambda_{\rm SO}$, which gives an effective magnetic field depending on spin and also an excitation gap $\Delta=3\sqrt{3} \lambda_{\rm SO}$ 
to the electron\cite{KM2}. Thus, the term yields quantization of the spin Hall conductivity (SHC)\cite{KM1,KM2}, 
\begin{eqnarray}
\sigma_{xy}^s=\frac{e}{2\pi d}\frac{\Delta}{|\Delta|},   
\label{SHC}
\end{eqnarray}
where $d$ is the interlayer distance. 
The model can also have the Rashba extrinsic SO coupling $\lambda_R$, which 
breaks the inversion symmetry and is induced by an electric field perpendicular 
to the honeycomb lattice  plane. The term also breaks the conservation of electron spin $s_z/2$. 
We consider $\lambda_R=0$ case, first.  

We add the on-site Coulomb repulsion $U$. The Hamiltonian per layer is, 
\begin{eqnarray}
H&=&t\sum_{<ij>}c^\dagger_i c_j + i \lambda_{\rm SO} \sum_{<<ij>>}\nu_{ij} c^{\dagger}_i s_z c_j 
\nonumber\\
&&+ U\sum_i n_{i \uparrow}n_{i \downarrow},  
\label{latticeH}
\end{eqnarray}
where $c_i$ ($c^\dagger_i$) is the annihilation (creation) operator of an electron with spin at the $i$-th site and $t$ is the nearest neighbor hopping. The second term is the intrinsic SO term consisting of the next nearest neighbor hopping, and 
$\nu_{ij}=\frac{2}{\sqrt{3}} (\hat{\bm d}_1\times \hat{\bm d}_2)_z=\pm1$, where  $\hat{\bm d}_1$ and  $\hat{\bm d}_2$ are unit vectors along the two bonds 
where the electron moving from site $j$ to $i$ passes. We assume that $U$ is lower than the critical value $U_c$, above which the system tends toward 
the other insulating phase.\cite{Rachel-LeHur} Namely, we consider the phase which is continuously connected to the quantum spin Hall state $U=0$. 

Let us discuss how to deal with the electron correlation $U$. 
On-site coulomb repulsion can be written by the on-site spin-spin interaction; 
\begin{eqnarray}
U n_{i\uparrow}n_{i\downarrow}=\frac{U}{2}(n_{i\uparrow}+n_{i\downarrow}) - \frac{U}{6}(c^\dagger_i \vec{s} c_i)^2. 
\end{eqnarray}
The first term just gives the renormalization for the chemical potential and can be neglected. 
We introduce the auxiliary field $\vec{\varphi}_i$, which is a three-component vector in the spin space, 
and use the Stratonovich-Hubbard transformation \cite{Mahan} 
$$H \rightarrow H_{SH}=H+\Delta H,$$where, 
\begin{eqnarray}
\Delta H&=&\frac{U}{6}\sum_i (c^\dagger_i \vec{s} c_i - \frac{3}{2U} \vec{\varphi}_i)^2, 
\label{SHgauss}
\end{eqnarray}
and therefore, 
\begin{eqnarray}
H_{SH}&=&t\sum_{<ij>}c^\dagger_i c_j + i \lambda_{\rm SO} \sum_{<<ij>>}\nu_{ij} c^{\dagger}_i s_z c_j 
\nonumber\\
&&-\sum_i \vec{\varphi}_i \cdot c_i^\dagger \frac{\vec{s}}{2} c_i + \frac{3}{8U}\sum_i |\vec{\varphi}_i|^2. 
\label{latticeHSH}
\end{eqnarray}
The spin-spin interaction is eliminated in appearance, but we have a coupling between $\vec{\varphi}_i$ and  the electron spin, and a quadratic term of $\vec{\varphi}_i$ instead.

We consider the continuum limit and take into account the low energy electronic 
excitations around $K$ and $K^\prime$ points in the Brillouin Zone \cite{KM1,KM2}. 
We introduce the electromagnetic $U(1)$ gauge field \cite{note1} $A_\mu$ and $SU(2)$ spin gauge field $\vec{a}_\mu$ via 
covariant derivative, 
\begin{eqnarray}
i D_\mu= i \partial_\mu - e A_\mu + \vec{a}_\mu \cdot \frac{\vec{s}}{2},  
\end{eqnarray}
where $\vec{a}_0=\vec{\varphi}$ (the auxiliary field in the continuum limit) and $\vec{\bm a}$ is a constant external field introduced artificially to estimate the spin current. 
We define a parameter 
\begin{eqnarray}
g=\frac{4 U a^2 d}{3} >0, 
\end{eqnarray}
where $a$ is the lattice constant, and the microscopic Lagrangian density is\cite{note2},  
\begin{eqnarray}
{\cal{L}}&=&\Psi^\dagger \left\{i D_0 - iv (D_x \tau_z \sigma_x + D_y \sigma_y) 
+ \Delta \tau_z \sigma_z s_z \right\}\Psi
\nonumber\\
&&
+ \frac{\epsilon_0 E^2}{2}- \frac{B^2}{2\mu_0} -\frac{1}{2g}|\vec{a}_0|^{2},   
\label{microL}
\end{eqnarray}
where $\Psi=\Psi_{\tau\sigma s}$ is the eight-component Fermion field which is labeled by the eigenvalues of the diagonal components of valley spin $\vec{\tau}$, sublattice spin $\vec{\sigma}$ and real spin $\vec{s}/2$. $\epsilon_0$ and $\mu_0$ denote the dielectric constant and magnetic permeability, respectively.  
Note that, except for the last term, the Lagrangian (\ref{microL}) possesses the $U(1)_{\rm em} \times U(1)_z$ local gauge symmetry.  
The $SU(2)$ gauge symmetry is broken down to $U(1)_z$, since the SO term contains $s_z$. 

The calculation which will be shown below is quite similar to one presented in Ref.\cite{GMI}, although the physical meaning of the spin gauge field is different. We integrate out $\Psi$ and obtain the one-loop effective Lagrangian for the gauge fields in the low energy and long wavelength region. 
The result is\cite{GMI},  
\begin{eqnarray}
{\cal{L}}_{\rm eff}&=&-\frac{1}{2g}a_0^{z2}+{\cal{L}}_{\rm ind}, 
\label{Leff}
\\
{\cal{L}}_{\rm ind}&=&\sigma_{xy}^s \epsilon^{\mu\rho\nu} a_\mu^z \partial_\rho A_\nu 
+ \frac{\epsilon E^2}{2} - \frac{B^2}{2\mu}+\frac{\delta \epsilon}{2}({\bm \nabla} a_0^z)^2 
\nonumber\\
&&+ ({\rm terms~independent~of}~a_\mu^z~{\rm and}~A_\mu),  
\label{Lind}
\end{eqnarray}
where ${\cal{L}}_{\rm ind}$ stands for the induced part of the effective Lagrangian. The first term in Eq. (\ref{Lind}) is the BF term\cite{BF}. The coefficient is the quantized SHC given in Eq. (\ref{SHC}). Note that only $a_\mu^z$ couples to the electromagnetic gauge fields. This comes 
from the fact that the $SU$(2) symmetry is broken down to $U(1)_z$ symmetry by the SO coupling. Maxwell term is renormalized as\cite{Semenoff-Sodano-Wu,GMI}
\begin{eqnarray}
\epsilon&=&\epsilon_0+\delta \epsilon
\nonumber\\
&=&\epsilon_0 + \frac{e^2}{6 \pi |\Delta| d}, 
\label{epsilon}\\
\frac{1}{\mu}&=&\frac{1}{\mu_0} + \frac{e^2 v^2}{6 \pi |\Delta| d}. 
\label{mu}
\end{eqnarray}
The elastic term for  $a_0^z(=\varphi^z)$ is also induced. 
We can recognize that any potential terms (i.e., zeroth-order terms with respect to the derivative $\partial_\mu$) of $A_\mu$ and also $a_\mu^z$ in ${\cal{L}}_{\rm ind}$ 
are absent because of $U(1)_{\rm em} \times U(1)_z$ gauge symmetry in the Fermionic part of microscopic Lagrangian (\ref{microL}). 
Thus, the low energy and long wavelength physics of $A_\mu$ and $a_\mu^z$ 
is described definitely by Eq. (\ref{Leff}). 

The solution of the static equation of motion obtained from the Lagrangian (\ref{Leff}) has been investigated \cite{Goryo-Maeda-short,Goryo-Maeda}. 
The argument indicates that when the dimensionless parameter satisfies
\begin{eqnarray}
\sigma_{xy}^{s2}\mu g>1,
\label{condition}
\end{eqnarray}  
we obtain 
the perfect diamagnetism as the energetically favorable solution. Namely, we obtain 
\begin{eqnarray}
B(x)&=&B(0)\frac{\mu}{\mu_0} e^{-\kappa_0 x},  
\label{Meissner}
\end{eqnarray}
where, 
\begin{eqnarray}
\kappa_0^{-1}&=&\frac{1}{2 e \sigma_{xy}^s}\sqrt{\frac{\delta \epsilon}{\mu}\frac{\sigma_{xy}^{s2}\mu g}{\sigma_{xy}^{s2}\mu g-1}}. 
\end{eqnarray}
shows the penetration depth. 
We also obtain 
\begin{eqnarray}
a^z_0(x)&=&\frac{B(0)}{\sigma_{xy}^s \mu_0} e^{-\kappa_0 x}. 
\end{eqnarray}
We can see from the BF term in eq. (\ref{Lind}) that electric current flows perpendicular to 
the gradient of $a_0(x)$ \cite{Goryo-Maeda}, and this current screens the applied magnetic field.

Let us take into account the Rashba coupling.\cite{KM1,KM2} We add   
\begin{equation}
-\frac{\lambda_R}{2} \Psi^\dagger (\tau_z \sigma_x s_y - \sigma_y s_x) \Psi,  
\label{Rashba} 
\end{equation}
to the microscopic Lagrangian (\ref{microL}). Since the Rashba term breaks $U(1)_z$ symmetry, the symmetry-breaking terms are induced additionally in the effective Lagrangian Eq. (\ref{Leff}). Up to the Gaussian approximation, the important alteration is the 
quadratic term of $a_0^z$; 
\begin{eqnarray}
-\frac{1}{2g}a_0^{z2} \rightarrow \frac{1}{2} \left(m_a^2 - \frac{1}{g}\right) a_ 0^{z2},  
\label{replace}
\end{eqnarray}
where \cite{GMI}
\begin{equation}
m_a^2=\frac{\lambda_R^2}{6 \pi |\Delta| d}+{\cal{O}}(\lambda_R^3). 
\end{equation}
Eq. (\ref{replace}) indicates that the parameter $g$ is renormalized as 
\begin{eqnarray}
g \rightarrow \frac{g}{1-g m_a^2}. 
\label{replace2}
\end{eqnarray}
The other parameters $\sigma_{xy}^{s}$, $\epsilon$ and $\mu$ in Eq. (\ref{Leff}) also receive renormalization of the order ${\cal{O}}(\lambda_R^2)$\cite{GMI}. The essential point is that Eq. (\ref{condition}), which is the 
condition for the Meissner effect, is changed as (see eq. (\ref{replace2})) 
\begin{eqnarray}
\frac{\sigma_{xy}^{s2}\mu g}{1-g m_a^2}>1. 
\label{condition2}
\end{eqnarray}
Compared with the condition (\ref{condition}), Eq. (\ref{condition2}) 
becomes more attainable. $g m_a^2$ is  positive and proportional to $\lambda_R^2$, which can be changed by the electric field perpendicular to 
the two-dimensional plane.



The authors are grateful to N. Hatano, D. S. Hirashima, K.-I. Imura, S. Kurihara, T. Oka, M. Sato and M. Sato  
for their fruitful discussions and stimulated comments. 
J.G. is financially supported by Grant-in-Aid for Scientific Research 
from Japan Society for the Promotion of Science under Grant 
No. 18540381, and also supported by Core Research for Evolutional
Science and Technology (CREST) of Japan Science and Technology Agency.


\begin{thebibliography}{99}
\bibitem{note1}
We suppose that the electromagnetic field is also two-dimensional.
\bibitem{note2}
Hereafter, we use the Lagrange formalism. The reason is that, the BF term, which will be induced later 
and plays the crucial role in our discussion, does not appear in the Hamilton formalism \cite{BF}. 
The effects which come from the BF term is expressed by the constraint and the formulation becomes 
complicated in the Hamilton formalism.
\end{thebibliography}
\end{document}